\documentclass[12pt]{iopart}
\usepackage{amssymb}
\usepackage{txfonts}

\usepackage{iopams}
\usepackage{graphics}
\newcommand{\beq}{\begin{equation}}
\newcommand{\eeq}{\end{equation}}
\newcommand{\beqa}{\begin{eqnarray}}
\newcommand{\eeqa}{\end{eqnarray}}

\begin{document}

\title[Double-periodic quasi-periodic graphene superlattice...]{Double-periodic quasi-periodic graphene superlattice: non-Bragg band gap and electronic transport}

\author{Xi Chen$^{1,2}$\footnote{Corresponding author. xchen@shu.edu.cn}, Pei-Liang Zhao$^{1}$, and Qi-Biao Zhu$^{1}$}

\address{$^{1}$ Department of Physics, Shanghai University, 200444 Shanghai, China}
\address{$^{2}$ Departamento de Qu\'{\i}mica-F\'{\i}sica, UPV-EHU, Apdo 644, 48080 Bilbao, Spain}

\begin{abstract}
Electronic band gap and transport in quasi-periodic graphene superlattice of double-periodic sequence have been investigated.
It is found that such quasi-periodic structure can possess a zero-averaged wave number (zero-$\bar{k}$) gap which associated with an unusual Dirac point.
Different from Bragg gap, the zero-$\bar{k}$ gap is less sensitive to the incidence angle, and robust against the lattice constants.
The locations of Dirac point and multi-Dirac-points in the graphene superlattices of various sequences are also compared.
The control of electron transport over the zero-$\bar{k}$ band gap in graphene superlattice may facilitate the development of many graphene-based electronics.

\end{abstract}

\submitto{\JPD}
\maketitle

\section{Introduction}
Since the discovery of graphene in 2004, electronic band gap and transport in graphene have attracted considerable attention
because of the intriguing physics as well applications in graphene-based nanoelectronics \cite{Novoselov2004,Novoselov2005,Zhang-TS,Castro}.
Graphene has a unique band structure with the conductance and valance bands touching at Dirac point,
which leads to many unusual properties of transport \cite{Castro},
like half-integer quantum Hall effect, Klein tunneling, and the minimal conductance.
So far, the transport properties including Klein tunneling and resonant tunneling have been extensively investigated in various graphene-based heterostructures ranging from
single barrier \cite{Katsnelson-NG,ChenAPL}  to superlattice with electrostatic potential and magnetic barriers \cite{Bai,Barbier2009,Brey,Park2008,Park2009,Wang2010,YuXian,Abedpour,Bliokh,Zhao2011,Ma2012}.

Several works have been devoted to the new Dirac point in the band structure of graphene superlattice (GSL) \cite{Barbier2009,Brey,Wang2010,Zhao2011,Ma2012}, since such new DP, associated with zero-averaged wave number (zero-$\bar{k}$) gap, is of benefit to the controllability of electronic transport. Different from Bragg gaps, the zero-$\bar{k}$ gap emerged with new Dirac point is robust against the lattice constant, structural disorder \cite{Wang2010}, and external magnetic field \cite{YuXian}.
Besides, the structure ordering of quasi-periodic GSL is between
periodic and disordered systems, which could give rise to many interesting and significant phenomena, for example, fractal spectrum and self-similar behavior \cite{Mukhopadhyay,Sena}.
Recently, we have studied the zero-$\bar{k}$ gap and electronic transport in Fibonacci quasi-periodic GSL \cite{Zhao2011}.
Shortly afterwards, Ma et al. \cite{Ma2012} found that the zero-$\bar{k}$ gap also exists in other quasi-periodic GSL of Thus-Morse sequence.

As we know, the main examples of quasi-periodic models are the Fibonacci (FB), Thue-Morse (TM) and Double-Periodic (DP) structures \cite{Enrique}.
These quasi-periodic superlattice can be generated by the following substitution rules: $A \rightarrow AB$ and $B \rightarrow A$ for FB,
$A \rightarrow AB$ and $B \rightarrow BA$ for TM, and $A \rightarrow AB$ and $B \rightarrow AA$ for DP, where $A$ and $B$ are two different barriers in
the quasi-periodic superlattice. The DP quasi-periodic superlattice is obviously different from FB and TM ones. But the number of barrier layers
in DP sequence increases as $N = 2^n$ (like the TM sequence), where $n$ indicates the iteration order, and barriers B appear always isolated (like the FB sequence) \cite{Enrique}.
The electronic properties of FB, TM and DP lattice are of significance \cite{Luck}.
So it is worthwhile to study the electronic band gap and transport in DP GSL and make comparisons with different quasi-periodic systems.
Combing the results in FB and TM GSL \cite{Zhao2011,Ma2012},
the controllable conductance and shot noise in DP GSL may result in the practical applications of electron transport
in quasi-periodic GSL.

\section{Theoretical Model}

Consider a graphene-based DP quasi-periodic superlattice which is arranged by DP substitution rule, $A \rightarrow AB$ and $B \rightarrow AA$, where $A$ and $B$ are two different barriers,
the parameters are denoted by the potential barrier height $V_{A}$ and width $d_A$, the potential well height $V_B$ and width $d_B$.
For an $n$-th DP sequence, $S_n$, it contains elements $A$ and $B$,
and follows the relation, $S_n=S_{n-1}S^{\dag}_{n-1}$, $S^{\dag}_n=S_{n-1}S_{n-1}$ (for $n>1$) with
$S_1=AB$ and $S^{\dag}_1=AA$, which leads to
$S_3= ABAA$, $S_3= ABAAABAB$, $S_4=ABAAABABABAAABAA$ and so on.
The Hamiltonian of change carriers near the $K$ point inside a monolayer graphene is given by,
$
\hat{H}=-i\hbar v_{F} {\vec{\sigma} \cdot \vec{\nabla} }+V(x),
$
where the Fermi velocity $v_{F}\approx 10^{6}$m/s, and $\vec{\sigma}%
=(\sigma _{x},\sigma _{y})$ are the Pauli matrices and $V(x)$ denotes the potential barrier or well.
The solution of $\hat{H}$, acting on the electronic pseudospin wave functions, results in
the transfer matrix \cite{Wang2010}:

\beq
\label{barrier}
M_j (\Delta x, E, k_y)=  \left(\begin{array}{cc}
      \frac{\cos(q_j \Delta x - \theta_j)}{\cos \theta_j} & i \frac{\sin(q_j \Delta x)}{\cos \theta_j}
      \\
      i \frac{\sin(q_j \Delta x)}{\cos \theta_j} & \frac{\cos(q_j \Delta x + \theta_j)}{\cos \theta_j}
   \end{array}\right),
\eeq
which connects the wave functions at $x$ and $x+\Delta x$ inside the $j$th potential with $\theta_j = \arcsin(k_y/k_j)$,
where $k_j=(E-V_j)/\hbar v_F$, $k_{y}$ and $q_j$ are the $y$ and $x$ components of wavevector,
$q_{j}= \mbox{sign}(k_j) ({k^2_{j}-k_{y}^{2}})^{1/2}$ for $ k^2_{j}> k^2_{y}$,
otherwise $q_{j}=i  (k^2_{y}-k_{j}^{2})^{1/2}$. Consequently, the transmission coefficient $t=t(E,k_y)$ is found to be
\beq
t =\frac{2 \cos \theta_0}{(m_{22} e^{- i\theta_0}+m_{11} e^{i\theta_t})-m_{12} e^{ i(\theta_t-\theta_0)}-m_{21}},
\eeq
where $\theta_0$ and $\theta_t$ are incidence and exit angles, and $m_{ij} (i,j=1,2)$ is
the matrix element of total transfer matrix, $X[S_n]= \prod^{N}_{j=1} M_j (d_j, E, k_y)$. 
Once the transmission coefficient is obtained, we can calculate the total conductance $G$ \cite{Datta} and the Fano factor $F$ \cite{Beenakker-PRL}
in terms of
$
G=G_{0}\int_{0}^{\pi/2} T \cos \theta_0
d \theta_0
$,
and
$
F= \int_{-\pi/2}^{\pi/2}T(1-T) \cos \theta_0
d \theta_0/ \int_{-\pi/2}^{\pi/2} T \cos \theta_0
d \theta_0
$,
where $T=|t|^2$ and $G_0 = 2 e^2 m v_F L_y/\hbar^2$, with $L_y$ is the width of the graphene stripe in the $y$ direction.

\begin{figure}[]
\begin{center}
\scalebox{0.30}[0.30]{\includegraphics{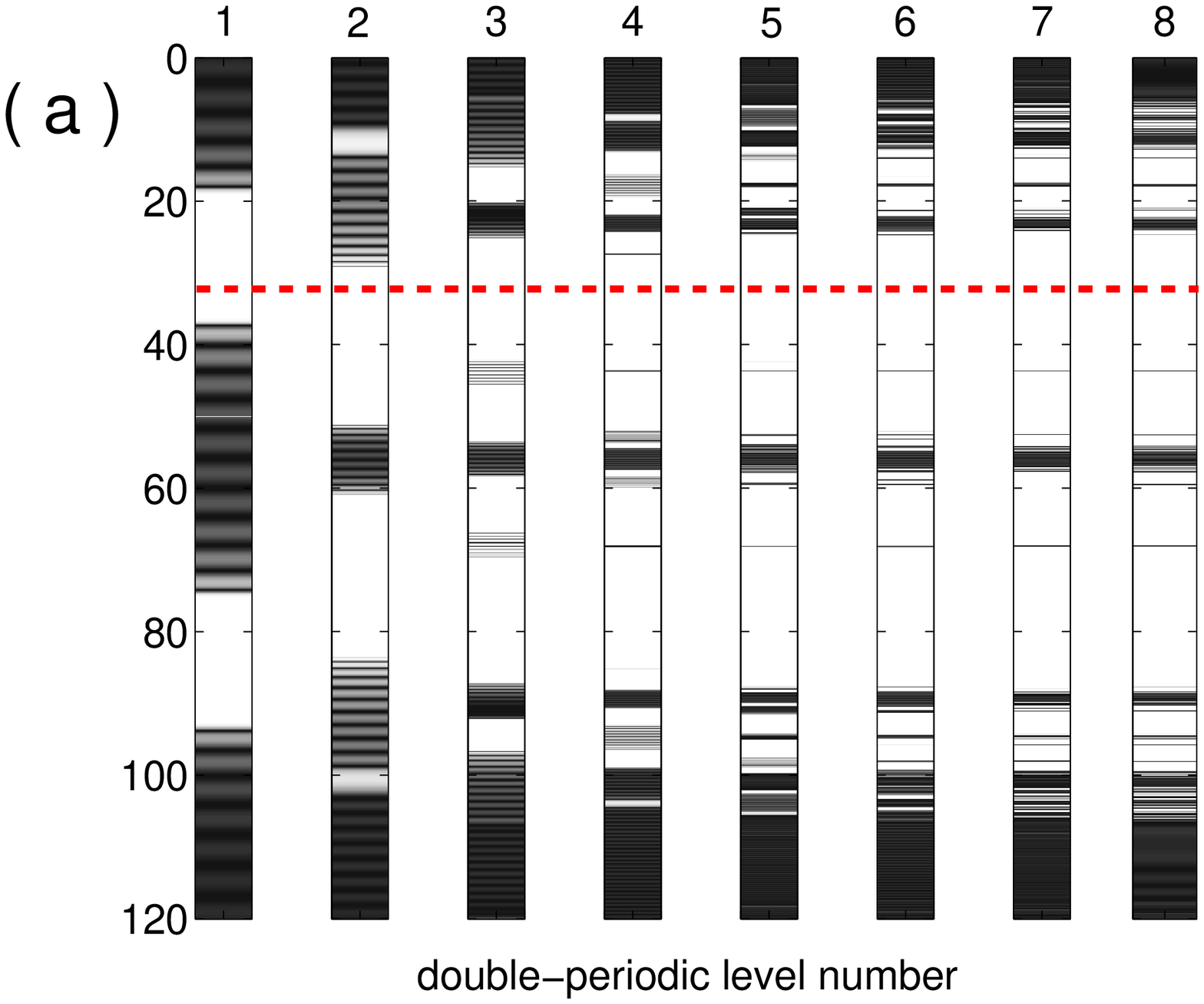}}
\scalebox{0.30}[0.30]{\includegraphics{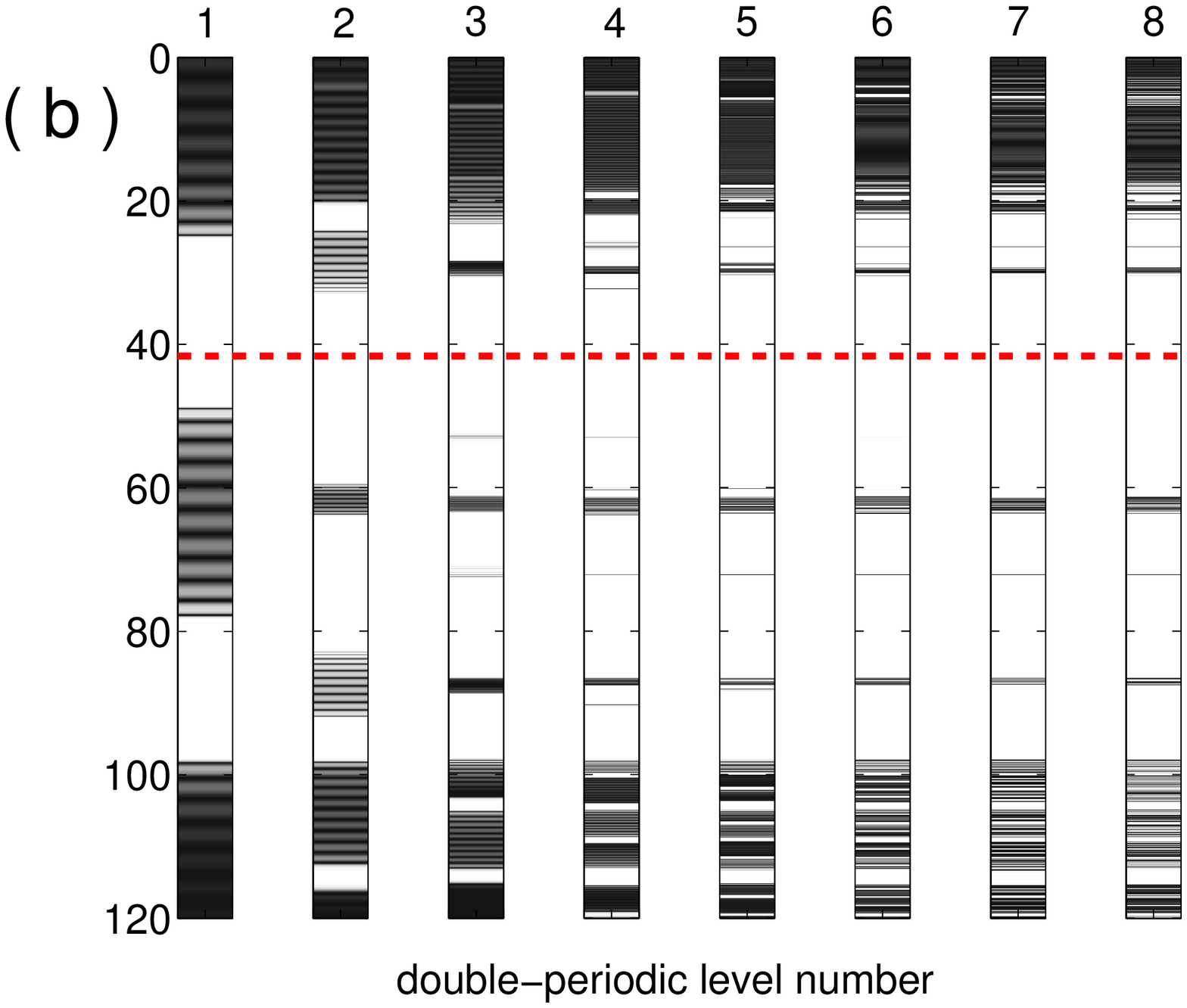}}
\caption{(Color online) Trace-maps for DP quasi-periodic GSL, $(S_n)^m$,
of variable order $n$ under (a) $d_A=d_B=20$ nm, (b) $d_A=2d_B=30$ nm,
where other parameters are $V_A=50$ meV, $V_B=0$ meV, $\theta_0=20^{\circ}$, and $m=12$.
The horizontal dashed red line denotes the new Dirac point's location. }
\label{fig.1}
\end{center}
\end{figure}

\begin{figure}[]
\begin{center}
\scalebox{0.65}[0.65]{\includegraphics{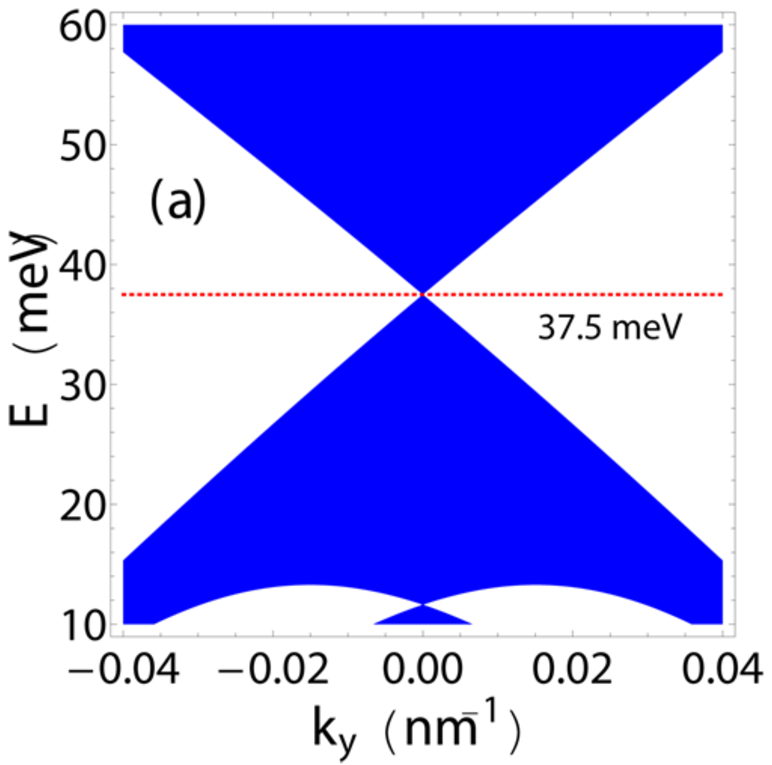}}
\scalebox{0.65}[0.65]{\includegraphics{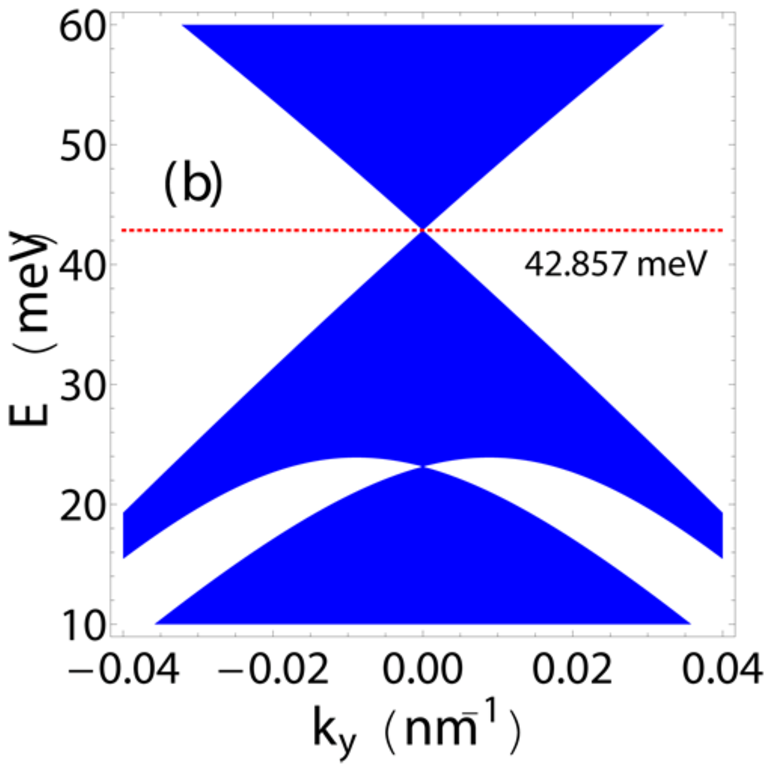}}
\caption{(Color online) Band structures for DP quasi-periodic GSL, $(ABAA)^{12}$, where (a) $d_A=d_B=20$ nm,
(b) $d_A=2 d_B=30$ nm, and the other parameters are the same as those in Fig. \ref{fig.1}.
The horizontal dashed red line denotes the new Dirac point's location.}
\label{dispersion}
\end{center}
\end{figure}
\begin{figure}[]
\begin{center}
\scalebox{0.6}[0.6]{\includegraphics{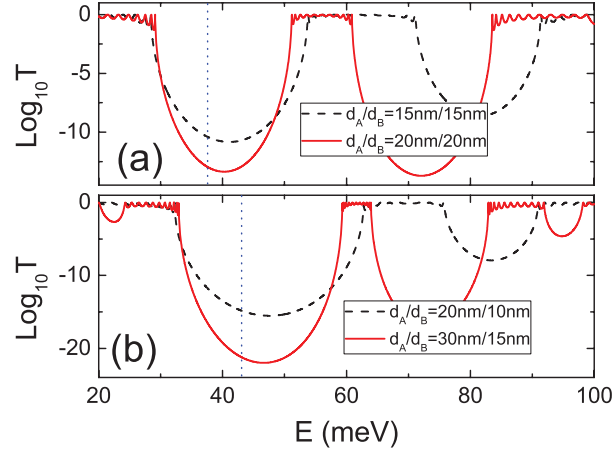}}
\caption{(Color online) Transmission spectra for DP quasi-periodic GSL, $(ABAA)^{12}$,
where (a): $d_A/d_B=1$, $d_{A}=15$ nm (dashed black line), $d_{A}=20$ nm (solid red line),
(b) $d_A/d_B=2$, $d_{A}=20$ nm (dashed black line), $d_{A}=30$ nm (solid red line),
and the other parameters are the same as those in Fig. \ref{fig.1}. }
\label{fig.3}
\end{center}
\end{figure}

In Fig. \ref{fig.1} (a) and (b), we plot the trace maps for quasi-periodic GSL of DP sequences, $S_n^m$, where (a) $d_{A}=d_B=20$ nm,
(b) $d_A=2d_B=30$ nm, the incidence angle $\theta_0=20^{\circ}$, $V_A=50$ meV, $V_B=0$ meV, and $m=12$.
Similar to the results for FB and TM sequences \cite{Zhao2011,Ma2012}, we find that there are several broad forbidden gaps open for each DP level,
and the passing bands are split into more and more subbands as DP sequence order $n$ increases.
Among these forbidden gaps, the center position of the zero-$\bar{k}$ gaps, which are denoted by a dashed red line,
are almost the same for different DP sequences in Fig. \ref{fig.1} (a) and (b).
These zero-$\bar{k}$ gaps are different from Bragg gaps, and thus are robust against the
lattice parameters.

To understand better, we further calculate the electronic dispersion, for example the second DP sequence ($S_2=ABAA$), at any incidence angle
\begin{equation}
\cos(\beta_x \Lambda_2)=\frac{1}{2} \mathbf{Tr} \{M_{A}M_{B}M_{A}M_{A}\},
\label{Bloch}
\end{equation}
where $\beta_x$ is the $x$ component of Bloch wave vector, and $\Lambda_2= 3 d_A + d_B$ is the length of the unit cell.
From Eq. (\ref{Bloch}), we can finally get the electronic band structures, as shown in Fig. \ref{dispersion} (a) and (b),
which correspond to (a) $d_A=d_B=20$ nm, (b) $d_A=2 d_B=30$ nm.
In general, the location of new Dirac point inside the zero-$\bar{k}$ gap is determined by
\begin{equation}
\label{condition}
\overline{k}=\sum\limits_{j=1}^{N} k_{j} d_{j}/\sum\limits_{j=1}^{N} d_{j}=0,
\end{equation}%
which is valid for both periodic and aperiodic GSLs \cite{Brey,Wang2010,Zhao2011,Ma2012}. Thus we achieve the new Dirac point
located at
\begin{equation}
E_{n}=\frac{V_{A}+V_{B}\cdot d_{B}/(\tau_{n}d_{A})}{1+d_{B}/(\tau_{n}d_{A})},
\label{Energy}
\end{equation}%
\begin{figure}[]
\begin{center}
\scalebox{0.6}[0.6]{\includegraphics{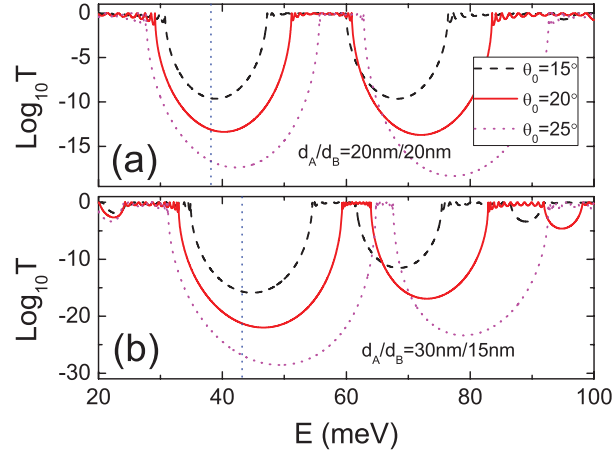}}
\caption{(Color online) Transmission spectra for DP quasi-periodic GSL, $(ABAA)^{12}$,
where (a): $d_A=d_B=20$ nm, (b): $d_A=30$ nm, $d_B=15$ nm, $\theta_0=15^{\circ}$ (dashed black line), $\theta_0=20^{\circ}$ (solid red line),
$\theta_0=25^{\circ}$ (dotted magenta line), and the other parameters are the same as those in Fig. \ref{fig.1}.}
\label{fig.4}
\end{center}
\end{figure}
which depends on the ratio of numbers of layer $A$ and $B$, that is, $\tau_{n}=N_{A}/N_{B}$. Here we would like to
emphasize that the location of Dirac point is generally dependent of the order $n$. But when TM sequence is considered,
the location of Dirac point does not depend on the order $n$ because of $\tau_{n}=1$ \cite{Ma2012}.
The different behaviors will be compared later in the different GSLs of FB, TM and DP sequences.
Besides, we see the Dirac point also depends on the ratio $d_{A}/ d_{B}$. As shown in From Fig. \ref{dispersion} (a) and (b),
the position of Dirac point is shifted from $E= 37.5$ meV for $d_A/d_B=1$ to $E=42.875$ meV for $d_A/d_B=2$. Furthermore, Fig. \ref{fig.3}
shows the transmission spectra for DP quasi-periodic GSL, $(ABAA)^{12}$. It is demonstrated
that the zero-$\bar{k}$ gap is robust against the ratio of lattice parameters, $d_{A}/d_{B}$, while other Bragg gaps
are sensitive to lattice parameters. From the point of view of application, this provides the flexibility to
control the electron transport by adjusting the ratio of lattice parameters, $d_A/d_B$, in the various
sequences.

Next, the influences of the incidence angle, $\theta_0$, on the transmission spectra are also studied. Obviously,
the zero-$\bar{k}$ gap associated with Dirac point is insensitive to the incidence angle, as shown in Fig. \ref{fig.4}.
On the contrary, the Bragg gap shifts upward in energy as the incidence angle increases. Compared to
the weak dependence of zero-$\bar{k}$ gap on the incidence angles, the Bragg gaps change sensitively with respect to
incidence angle $\theta_0$.

\begin{figure}[]
\begin{center}
\scalebox{0.65}[0.65]{\includegraphics{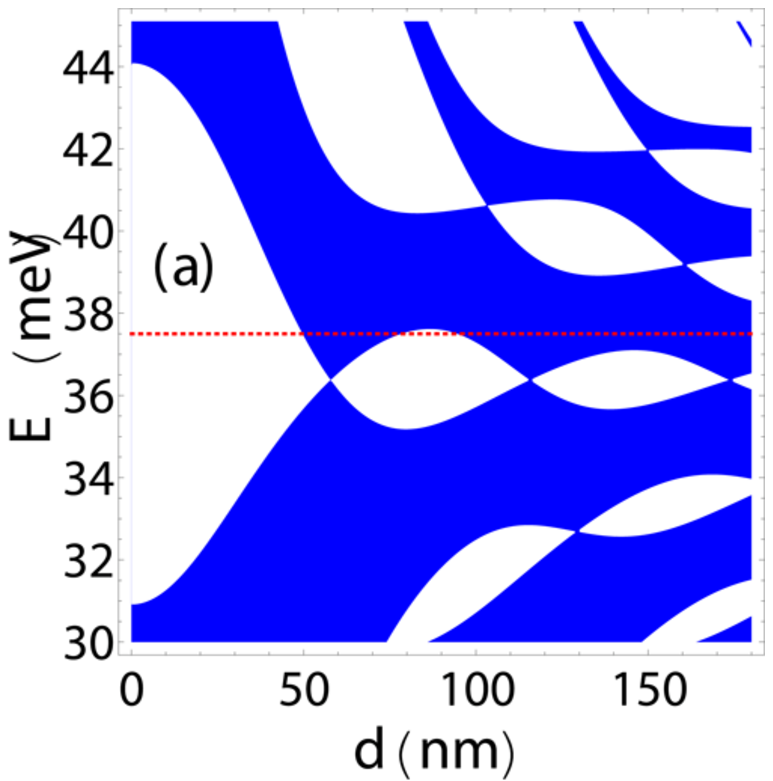}}
\scalebox{0.65}[0.65]{\includegraphics{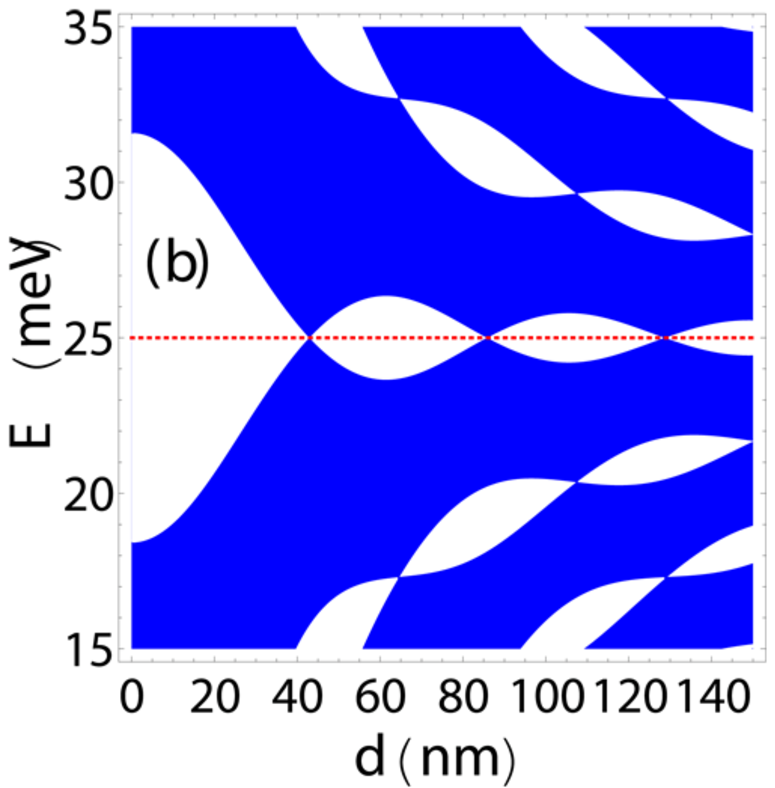}}
\caption{(Color online) Electronic band gaps depending on the lattice constants $d_A=d_B=d$ with $k_y = 0.01$ nm$^{-1}$, where
(a) $S_2= ABAA$ for DP sequence, (b) $S_2= ABBA$ for TM sequence, and the other parameters are the same as those in Fig. \ref{dispersion}.
The horizontal dashed red line denotes the Dirac point's location for $k_y=0$, where (a) $E=37.5$ meV and (b) $E=25$ meV. }
\label{fig.5}
\end{center}
\end{figure}

Additionally, the multi-Dirac-points could appear in the GSL with periodic \cite{Brey,Wang2010} and aperiodic \cite{Ma2012}
structures. Fig. \ref{fig.5} demonstrates that the extra Dirac points located at $k_y \neq 0$, can also emerge in the GSL of DP sequence.
We show the energy band gaps versus the lattice constants in the case of $d_A=d_B=d$.
The zero-$\bar{k}$ gap, associated with the multi-Dirac-points, oscillates with changing the lattice constants, thus open and close periodically,
while the other Bragg gaps are significantly shifted. However,
the multi-Dirac-points in DP sequence are different from those in TM sequence. To compare them, we choose $S_2= ABAA$ for DP and $S_2=ABBA$
for TM sequence. We see from Fig. \ref{fig.5} (a) that the multi-Dirac-points appear at the same energy, $E= 36.4$ meV,
which is different from the location, $E=37.5$ meV, of Dirac point for $k_y=0$.
Moreover, the energy corresponding the multi-Dirac-points in DP sequence will be close to $E=37.5$ meV, when $k_y \rightarrow 0$.
However, the Dirac point for $k_y=0$ and multi-Dirac-points for $k_y \neq 0$ happen at the same energy $E=25$ meV in TM sequence with $\tau_n=1$.
It tells us that though the total numbers of layer $A$ and $B$ are the same for TM and DP sequences, the behaviors of the
multi-Dirac-points in TM and DP sequences are totally different, because the ratio, $\tau_n = N_A/N_B$, are not equal.
\begin{figure}[]
\begin{center}
\scalebox{1.0}[1.0]{\includegraphics{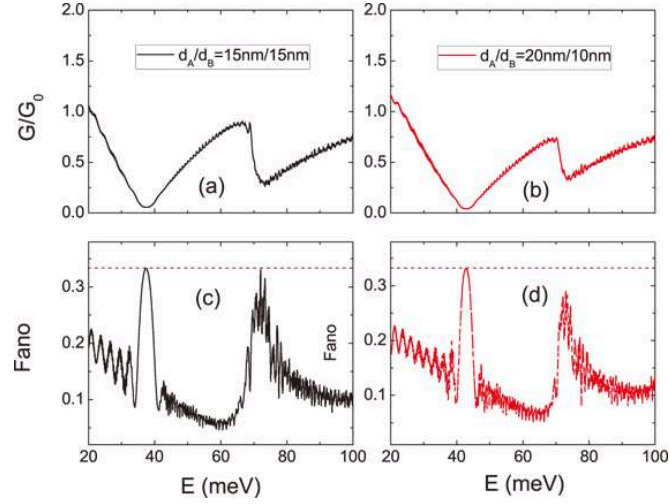}}
\caption{(Color online) Conductance $G/G_0$ [(a) and (b)] and Fano factor $F$ [(c) and (d)] versus Fermi energy $E$ in DP quasi-periodic GSL, $(ABAA)^{12}$,
where (a,c) $d_A/d_B =1$, $d_A=15$ nm, (b, d) $d_A/d_B =2$, $d_A=20$ nm, and the other parameters are the same as those in Fig. \ref{fig.1}.}
\label{fig.6}
\end{center}
\end{figure}

\begin{figure}[]
\begin{center}
\scalebox{0.50}[0.50]{\includegraphics{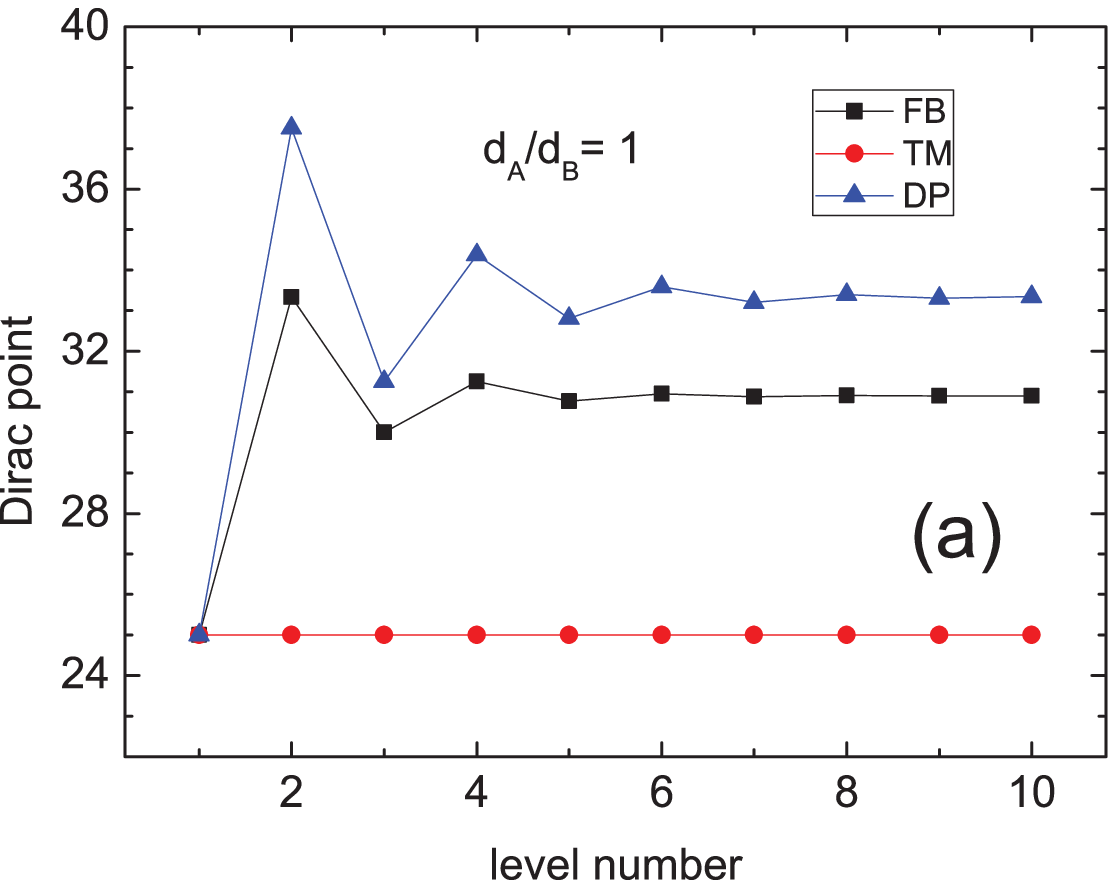}}
\scalebox{0.50}[0.50]{\includegraphics{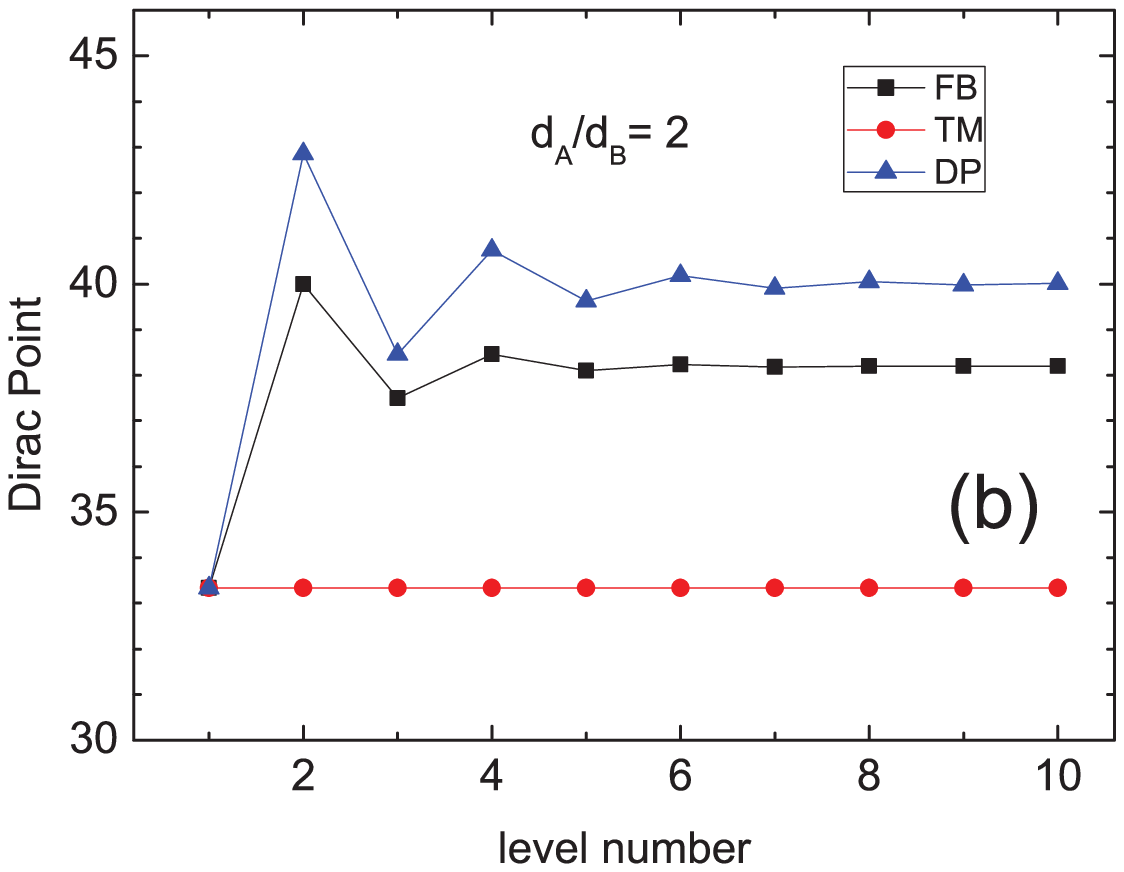}}
\caption{(Color online) Location of Dirac point versus variable order $n$ in FB, TM and DP sequences, where (a) $d_A/d_B =1$, (b) $d_A/d_B =2$,
and the other parameters are the same as those in Fig. \ref{fig.1}.}
\label{fig.7}
\end{center}
\end{figure}

Regarding the electronic transport, the total conductance and the Fano factor are also interesting. In Fig. \ref{fig.5}, we calculate
the total conductance $G/G_0$ and the Fano factor $F$ in quasi-periodic GSL of DP squence, $(ABAA)^{12}$. Again,
we have found serval unique features associated with new Dirac point and zero-$\bar{k}$ gap, as shown in Fig. \ref{fig.6}.
Firstly, the curve of angular-averaged conductance reaches its minimum at the Dirac point, and forms a liner cone around Dirac point.
Meanwhile the Fano factor reaches the value of $1/3$ approximately \cite{Beenakker-PRL}.
Secondly, the conductance and the Fano factor will shift with the ratio of $d_A/d_B$. Actually, this will be useful to
modulate the conductance of such GSL by changing the ratio of lattice constants.

Finally, what we should mention is that the transmission spectrum and corresponding electronic transport properties including
conductance and the Fano factor in the DP sequence are quite different from those in FB and TM sequences. To clarify it, we shall
compare the Dirac point in three different sequences. In Fig. \ref{fig.7}, we present the location of Dirac point versus
the variable order $n$ in FB, TM and DP sequences. In TM sequences, the location of new Dirac point does not depend on
the order $n$, because, the numbers of layer $A$ and $B$ are the same as mentioned above. As for FB and DP,
the numbers of layer $A$ and $B$ are different in each order $n$ ($n>1$). As a consequence, the location of Dirac point depends on
the variable order $n$, but it will becomes stable for $n \gg 1$, since the ratio of numbers of layer $A$ and $B$, $N_A/N_B$,
trends to $\tau_n = (1+\sqrt{5})/2$ for FB sequence \cite{Zhao2011,Enrique} and $\tau_n = 2$ roughly for DP sequence \cite{Enrique}.
These difference presented here are quite natural, because the physical origin of zero-$\bar{k}$ gap and new Dirac point is
the total zero phase, suggested by Eq. (\ref{condition}).


\section{Summary}

To summarize, we have studied the electronic band gap and transport in GSL of DP sequence using the transfer matrix method.
We have found that the Dirac point, multi-Dirac-points and associated zero-$\bar{k}$ gaps
have emerged in such quasi-periodic structure. The zero-$\bar{k}$ gap and Dirac point are robust against the lattice constants and incidence angle. The results are also
compared with those in GSL of FB and TM sequences.
We hope that all these results may lead to applications in the control of electron transport in GSL.

\ack{
We acknowledge funding by the National Natural Science
Foundation of China (Grant Nos. 60806041, and 61176118) and the Science and Technology Committee of Shanghai Municipality (Grant No. 11ZR1412300).
X. C. is also supported by the Basque government (Grant No. IT472-10), Ministerio de Ciencia e Innovacion (Grant No.
FIS2009-12773-C02-01), and the UPV/EHU under program
UFI 11/55.}

\end{document}